# Instability and Surface Potential Modulation of Self-Patterned (001)SrTiO$_3$ Surfaces

Lucía Aballe,[1] Sonia Matencio,[1] Michael Foerster,[1] Esther Barrena,[2] Florencio Sánchez,[2] Josep Fontcuberta,[2] and Carmen Ocal[2,*]

[1] ALBA Synchrotron Light Facility, Carretera BP 1413, Km. 3.3, 08290 Cerdanyola del Valles, Barcelona, Spain
[2] Institut de Ciència de Materials de Barcelona (ICMAB-CSIC), Campus UAB, 08193 Bellaterra, Barcelona, Spain

* Corresponding Author: cocal@icmab.es

ABSTRACT:
The (001)SrTiO$_3$ crystal surface can be engineered to display a self-organized pattern of well-separated and nearly pure single-terminated SrO and TiO$_2$ regions by high temperature annealing in oxidizing atmosphere. By using surface sensitive techniques we have obtained evidence of such surface chemical self-structuration in as-prepared crystals and unambiguously identified the local composition. The contact surface potential at regions initially consisting of majority single terminations (SrO and TiO$_2$) is determined to be $\Phi(SrO) < \Phi(TiO_2)$, in agreement with theoretical predictions, although the measured difference $\Delta\Phi \leq 100$ meV is definitely smaller than theoretical predictions for ideally pure single-terminated SrO and TiO$_2$ surfaces. These relative values are maintained if samples are annealed in UHV up to 200 °C. Annealing in UHV at higher temperature (400 °C) preserves the surface morphology of self-assembled TiO$_2$ and SrO rich regions, although a non-negligible chemical intermixing is observed. The most dramatic consequence is that the surface potential contrast is reversed. It thus follows that electronic and chemical properties of (001)SrTiO$_3$ surfaces, widely used in oxide thin film growth, can largely vary before growth starts in a manner strongly dependent on temperature and pressure conditions.

INTRODUCTION

Electronic redistributions at solid interfaces occur to reduce the chemical potential gradients, and the energy cost to bring electrons in/out across these interfaces is strongly influenced by the intrinsic properties of materials involved, namely, the work function ($\Phi$) in metals and the electron affinity ($\chi$) and the ionization potential (IP) in intrinsic semiconductors and insulators.

The work function of metals and degenerated semiconductors depends on the electron density and surface dipoles formed at the corresponding free surface. Similarly, the IP of insulators and intrinsic semiconductors depends on the crystallographic plane considered. Modulation of the substrate work function, as required in most common planar electronic devices or active surfaces, can be achieved by changing the carrier concentration by chemical or electrostatic doping, or by chemically engineering the surface. A beautiful example is the LaAlO$_3$−SrTiO$_3$ interface, where a high mobility two-dimensional electron gas (2DEG) can be formed only if the STO surface is TiO$_2$ terminated [1]. Although, following the pioneering work by Kawasaki et al. [2], TiO$_2$ single-terminated (001)SrTiO$_3$ (STO) single-crystalline surfaces can be prepared, as-received crystals display

SrO and TiO$_2$ terminations. Therefore, it follows that, in a surface with chemically separated but coexisting TiO$_2$ and SrO regions, highly conducting 2DEG patches of nanometric lateral size can be embedded in an insulating matrix, [3] thus opening the possibility to explore the physics of confined 2DEGs.

Chemical self-assembling at surfaces of ABO$_3$ perovskites with (001) orientation has been proposed to be a suitable technological approach for lateral surface nanostructuration [4]. It has been shown that the two distinct terminations AO and BO$_2$, which are randomly distributed in as received (001) single- crystalline substrates, can be promoted to order in well-defined regions having in majority either AO or BO2 composition, differing by ½ unit cell (uc) in height (≈0.2 nm).As reported in previous scanning probe microscopy studies concerning (001)STO structuration [5,6], such regions can form either confined leaf-shaped regions at the intermediate stages of self- separation or regularly ordered arrays of stripes, some few tens of nanometers wide and several hundred microns long, once the surface reordering is fully developed. A modulated electrostatic contact potential difference (CPD) was measured at ambient conditions by Kelvin probe force microscopy (KPFM), showing a pattern that replicated the self-ordered surface topography [7]. Thus, (001) surfaces composed by nanometric regions of SrO and TiO$_2$ terminations could be viewed as templates with modulated surface potential, where the potential contrast indicates the different Φ or IP of each chemical termination. Both CPD measurements on each termination of self-ordered surfaces and IP calculations of two ideal single-terminated surfaces gave larger Φ values for BO$_2$ than for AO [7]. However, disagreement between absolute values of experimental ΔCPD (≈45−70 mV) and ΔIP calculations (≈2 eV) left open questions related not only to the finite size of the nanostructuration and its convolution with measuring procedures (namely, tip size and tip−sample distance) but also to the role of surface adsorbates [8] and/or surface reconstructions in minimizing surface energy and reducing surface potential differences in the experiments when compared to the theoretical ones. In fact, the large energy difference between AO and BO$_2$ regions predicted by theory should create an unstable surface unless additional electronic reconstructions or adsorbate induced stabilization take place. Therefore, atomic rearrangement, species segregation, and/or diffusion are likely to occur [9]. Assignment of the distinct self-assembled surface regions of (001)STO to SrO and TiO$_2$ composition [5,6,10−12] is based on subtle morphological features or rule-of-thumb chemical considerations on stability and reactivity, or distinct friction forces in different regions. [13,14] However, using only scanning probe microscopy techniques which do not provide chemical details, it is not possible to univocally identify each termination.

In summary, two fundamental questions arise. First, what would be the ΔCPD values if measurements were performed under UHV conditions minimizing the role of adsorbates? Second, what is the actual chemical termination of the distinctive regions observed on the (001)STO surfaces? Determining the actual chemical and electrostatic properties of such nanostructured surfaces is particularly important since this and other closely related perovskite oxides are much used as substrates for the growth of epitaxial thin films, and hence surface potential modulations could imprint interesting modulated properties on the overlayers.

Aiming to identify with high lateral resolution the chemical nature of distinct self-arranged terminations and to map their corresponding work function, we present here a comprehensive study of selected nanostructured (001)STO surfaces by combining contact-mode atomic force microscopy (c-AFM) with other surface sensitive techniques in ultrahigh vacuum (UHV), namely, noncontact atomic force microscopy (nc-AFM), X-ray photoemission electron microscopy (XPEEM), and low energy electron microscopy (LEEM).

We have obtained photoelectron spectroscopy maps with submicron lateral resolution of the crystal surfaces allowing determination of the chemical nature of the distinct regions and univocally assigning them to majority SrO or TiO$_2$ terminations. Our results indicate that the work-function differences between these surface terminations can be obtained either in UHV or ambient conditions giving consistent results but are dramatically dependent on the surface preparation, in particular, on annealing conditions (atmosphere and temperature). It is found that, for crystals annealed at mild reducing conditions (up to 200 °C in UHV), the surface potential contrast is as in the as-prepared (oxygen-annealed) samples, i.e., SP(SrO) < SP(TiO$_2$) and ΔCPD ≈ (45−70) mV. However, even in these moderate annealing conditions some incipient chemical changes are observed. Moreover, we demonstrate that surface reconstruction and faint intermixing, resulting from subsequent thermal treatment at only 400 °C in UHV, have dramatic consequences on the surface potential map. Specifically, the sign of the surface potential difference between regions initially consisting of nominally pure single terminations (SrO and TiO$_2$) is reversed. Whereas instability of STO surface under UHV conditions at elevated temperatures had been already reported [15], the present results emphasize chemical and electrostatic implications of surface reconstruction signaling a narrower temperature window of surface stability.

EXPERIMENTAL

(001)STO single-crystalline substrates from CrysTec GmbH were thermally treated at high temperature (≈1150 °C) in air for 2 h in a dedicated tubular furnace: in the following these samples are labeled "as-prepared". Screening characterization of STO samples to select the desired chemical nanostructuration was done by c-AFM (Nanotec Electronica S.A.) combining topographic and friction force microscopy (FFM) [16] at room temperature in a N$_2$ environment with a relative humidity <5%. Freshly prepared samples were measured as transferred from the tubular furnace; if the process lasted more than 1 h, right before c-AFM measurements, the samples were cleaned with acetone and subsequently rinsed in ethanol. Silicon tips mounted on low spring constant cantilevers (k = 0.5 N/m) were employed.

An extended characterization was carried out in UHV by nc-AFM in frequency modulation with an amplitude of 200 pm, using standard Kolibri sensors with resonance frequency 1 MHz and k ∼ 540 kN/m (Specs GmbH). The CPD was determined from the parabolic dependence of the frequency shift (Δf) versus bias voltage (V) at specific surface points. In our UHV setup, the voltage bias is applied to the sample. Hence, larger CPD corresponds to larger local work function (see details on bias voltage sign convention in Supporting Information). All AFM data were analyzed by using the WSxM freeware [17]. The samples were cleaned with acetone plus ethanol before they were introduced into the UHV chamber, followed by 15 min of in situ thermal annealing. Three thermal treatments have been performed: at 75 °C (for which neither adsorbates desorption nor chemical changes are expected) and to 200 and 400 °C to reduce undesired adsorbates contribution and check the effects of annealing. Samples used in this study are coded as sample **1a** (as-prepared), sample **1b** (**1a** + 200 °C/ UHV), sample **1c** (**1b** + 400 °C/UHV), sample **2a** (as-prepared), and sample **2b** (**2a** + 75 °C/UHV).

XPEEM and LEEM were carried out at the spectro-microscope of the CIRCE beamline at the ALBA Synchrotron Light Facility in Barcelona (Spain). The beamline provides a highly focused beam of polarized photons in the soft X-ray range (100-2000 eV) and has a spectroscopic PEEM-LEEM instrument [18] on one of its branches, providing a variety of surface characterization techniques in a single instrument [19].

RESULTS AND DISCUSSION

As-received (001)STO single-crystalline substrates (see Supporting Information) have terraces separated by 1 uc steps (0.3905 nm) and exhibit a random distribution of SrO and $TiO_2$ regions of size below the lateral resolution of c-AFM (from now on "intermixed" surface). As described in detail elsewhere [5], by an appropriate thermal treatment, these two terminations are promoted to self-arrange, resulting in elongated patches confined between terraces (see below). If the surface thermal treatment is prolonged, these leaf-shaped structures evolve to eventually form alternated $TiO_2$ and SrO stripes differing by ½ uc in height, regularly ordered on the crystal surface. As the final surface details depend on specific sample characteristics, including wafer miscut, here different (001)STO single crystals were prepared following a similar thermal treatment (2 h at 1150 °C in air atmosphere) to obtain regions of majority SrO or $TiO_2$ terminations. In the present study, to make easier the identification of similar regions when comparing non-simultaneous measuring techniques, surfaces with clearly distinguishable morphologies (leaf-shaped or others) were preferred.

An illustrative example of the morphology of an as-prepared (001)STO surface (sample 1a) is shown in the c-AFM images of Figure 1a-c. Large terraces (labeled B) separated by confined regions (labeled A) about 100 nm wide are distinguishable. The height histogram (Figure 1b) from data in Figure 1a shows that the difference in altitude between B regions is about 0.4 nm, closely coinciding with 1 uc in STO, whereas the height difference between A and B is about 0.2 nm, corresponding to ½ uc. This observation indicates that these regions correspond to alternate (001) planes of STO having, predominantly, different atomic terminations, either SrO or $TiO_2$. Chemical separation is further supported by FFM data revealing a higher friction at A than at B regions, as evidenced in Figure 1c-e, where topographic c-AFM large-scan images (Figure 1c), and forward (Figure 1d) and backward (Figure 1e) lateral force images, are shown.

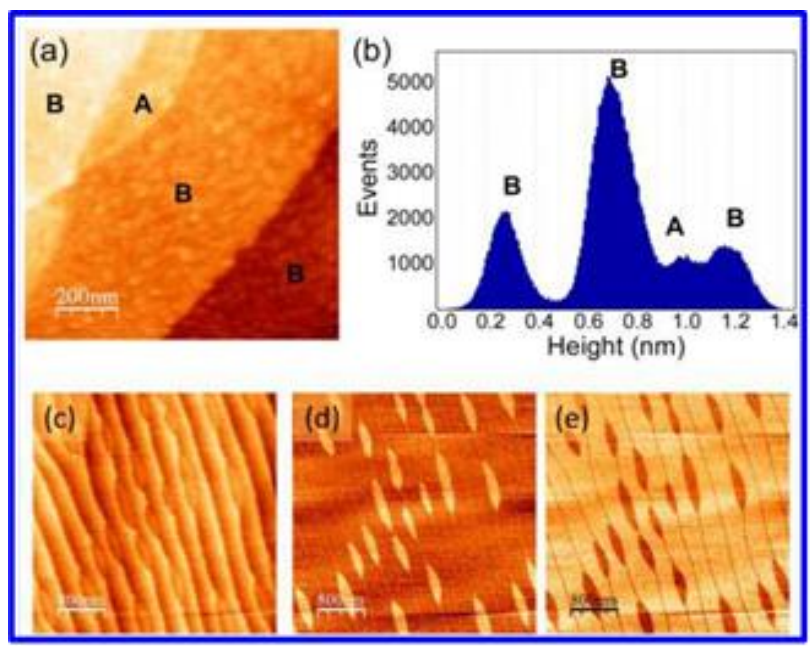

Figure 1. (a) Topographic image measured by c-AFM (sample 1a), in contact mode (under $N_2$ atmosphere) and the corresponding height histogram (b) showing that A leaf-shaped regions have an intermediate height (≈0.2 nm) between consecutive B terraces, which are separated by ≈0.4 nm. The topography and the corresponding forward and backward lateral force images of a larger region of the same sample are shown in parts c, d, and e, respectively. The contrast observed in parts d and e indicates a larger friction of the confined regions (A) as compared to the terraces (B).

As commented above, on the basis of indirect information, such as step-edge shape, and reactivity to moisture or selective growth, it has been proposed that A regions correspond to SrO and B regions to $TiO_2$. However, only on the basis of these data, this assignment cannot be unambiguously confirmed. In addition, within each region, the presence of minority fractions of the complementary termination cannot be excluded.

The distinct nature of the two terminations should also be reflected in the local electrostatic potential of the surface. In order to measure the CPD in UHV, in addition to the ex-situ cleaning protocol, the sample was treated in situ (base pressure $5 \times 10^{-10}$ mbar) at 200 °C during 15 min in order to desorb water and minimize adsorbates which might mask the true surface electrostatic potential [8] (sample 1b). In Figure 2a we show the corresponding nc-AFM topographic image, where well-differentiated regions, differing in altitude by ½ uc, reveal the presence of SrO and $TiO_2$ terminations at this surface. The CPD value at selected locations was determined from the position of the maximum of the parabolic dependence of $\Delta f(V)$. Figure 2d shows the values obtained at selected points on each region. The corresponding parabolic fits are also presented. The result indicates CPD(A) < CPD(B), i.e., $\Phi(A) < \Phi(B)$, with $\Delta\Phi$ ranging from +35 to +65 meV after measuring over 10 points at each region. These results match exceptionally well with those obtained in $N_2$ atmosphere at low humidity conditions in as-prepared samples (see Supporting Information). It is worth stressing that data collected either in as-prepared samples as sample 1a, and samples prepared by a mild low temperature annealing (200 °C) under UHV conditions (sample 1b), consistently indicate that the work function of A is smaller than that of B regions, in agreement with the theoretical trend for A corresponding to SrO and B to $TiO_2$ [7].

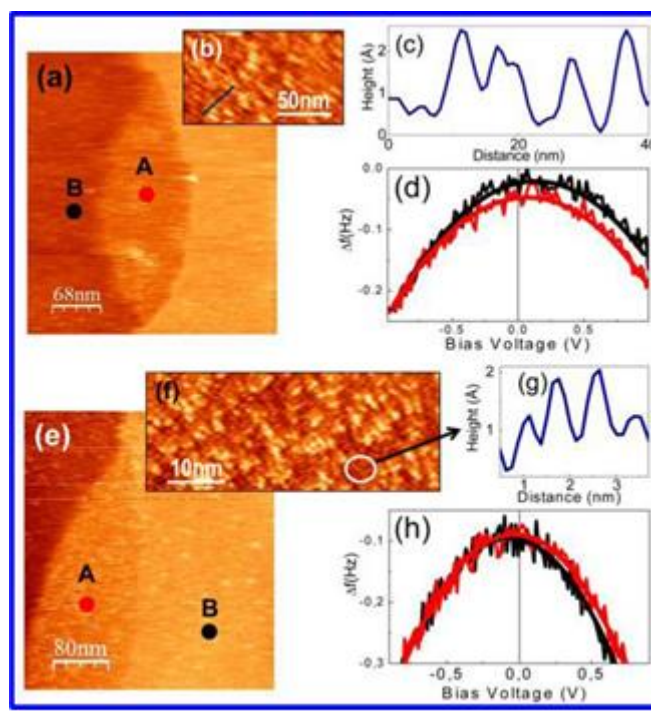

Figure 2. (a) Topographic image obtained by nc-AFM ($\Delta f$ = +795 mHz, V = −20 mV) after sample annealing at 200 °C during 15 min under UHV (sample 1b). (b) Magnified high resolution topography on the B terrace. (c) Topographic profile along the 40 nm long segment indicated in part b. (d) $\Delta f$ versus V spectroscopy on A (red) and B (black) regions. The maxima of the parabolic fits (continuous lines) give CPD (A) = +69 mV and CPD (B) = +134 mV. (e) Topographic nc-AFM image ($\Delta f$ = +320 mHz, V = −50 mV) acquired after sample annealing at 400 °C during 15 min under UHV (sample 1c). (f) Magnified high resolution topography on the B terrace. (g) Line profile (4 nm long) taken at the encircled region of part f. (h) $\Delta f(V)$ measured on B (black) and A (red) regions of part e. The CPD values obtained from the parabolic fits (continuous lines) are CPD (A) = −35 mV and CPD (B) = −67 mV.

However, high resolution nc-AFM images (Figure 2b) show some degree of heterogeneity with fairly periodic stripes of aligned protrusions ≈0.1−0.2 nm high and 10 nm wide (Figure 2c) which should not exist on ideally $TiO_2$ surfaces. This observation suggests an incipient partial mixture of terminations that deserves attention and is addressed in the following, where we characterize the surface of the same crystal after further UHV annealing at higher temperature (15 min 400 °C) (sample 1c).

Figure 2e is a topographic image of this high temperature annealed sample (sample 1c), and Figure 2f is a high resolution image acquired by nc-AFM on a B terrace. Although adsorbate desorption could occur at this temperature range, in a comparison of data from sample 1b and sample 1c (Figure 2b and Figure 2f) it is clear that the thermal treatment at higher temperature (400 °C) promotes surface reconstruction in which the protrusions, already seen in samples treated at 200 °C UHV, are no longer forming stripes but are randomly distributed. They have an apparently larger vertical size of ≈0.2−0.3 nm (see Supporting Information). Similar protrusions have been observed on top of reconstructed $TiO_2$ terminated (001)STO surfaces and ascribed to Sr or $SrO_x$ related clusters [20−22]. Accordingly, these features are considered here to be evidence of a defective $TiO_2$ surface with partially intermixed chemical composition, containing also SrO. Beside the brighter protrusions of Figure 2f, a relatively well-ordered surface lattice is visible (encircled region and corresponding profile in Figure 2g), which is compatible with the (√5 × √5) R26.6° (001)STO reconstruction previously reported by UHV- STM and nc-AFM data [21−25] and modeled as Sr adatoms trapped on the oxygen 4-fold site of the (001)$TiO_2$ termination [23]. In Figure 2h we show the Δf(V) obtained at two different points on regions A and B of sample 1c. Interestingly the Δf(V) parabola shift gives Φ(A) > Φ(B), which is opposite to that observed in Figure 2d (sample 1b) and contrary to the theoretical predictions if A and B were pure SrO and $TiO_2$, respectively [7]. The measured difference in CPD is such that ΔΦ ranges from −20 to −30 meV after measuring over 10 points at each region. The following discussion will demonstrate that Φ(A) < Φ(B) or Φ(A) > Φ(B) strongly depends on the degree of Sr surface segregation. To confirm that the sign of the measured work-function difference comes from the compositional details of each surface, thus highlighting the relevance of any unbalance in the chemical separation on the actual surface potential, we explored other ways of obtaining experimental work functions.

In principle, the work function can also be mapped with high lateral resolution by LEEM in the so-called mirror electron microscopy (MEM) mode. Here the sample is illuminated in normal incidence with very low energy electrons which are reflected in front of the surface and deflected by the field inhomogeneities, and then used to form a magnified image [19]. The electron-energy-dependent transition from the total reflection MEM mode to the LEEM regime, where the electrons penetrate and interact with the sample, is a measurement of the surface electrostatic potential or work function.

The MEM images of sample 1c collected at different electron energies close to the mentioned MEM−LEEM transition are shown in Figure 3a−d. The surface features observed by nc-AFM (Figure 2e) can be clearly recognized in MEM as darker elongated regions (A) surrounded by a lighter background (B regions). Starting from the left (Figure 3a,b), the A regions appear darker than the B regions (the electron reflectivity of the B termination is higher than that of A regions). With increasing electron energy, the contrast decreases until it practically vanishes which signals the onset of the LEEM regime (Figure 3c,d). At the very onset, however, there is a narrow energy window where the A termination appears at first nonhomogeneous in brightness, possibly due to the mentioned partial intermixing (Figure 3c), and eventually becomes brighter than B (Figure 3d).

The 1 uc high steps separating different terraces that are seen in the friction data of Figure 1 as fine lines are also clearly visible as fine dark straight lines in the MEM images of Figure 3. The observed contrast at step edges in lateral force FFM results from cantilever torsion when crossing steps connecting terraces at different levels; in MEM it results from a combination of topography and surface potential difference.

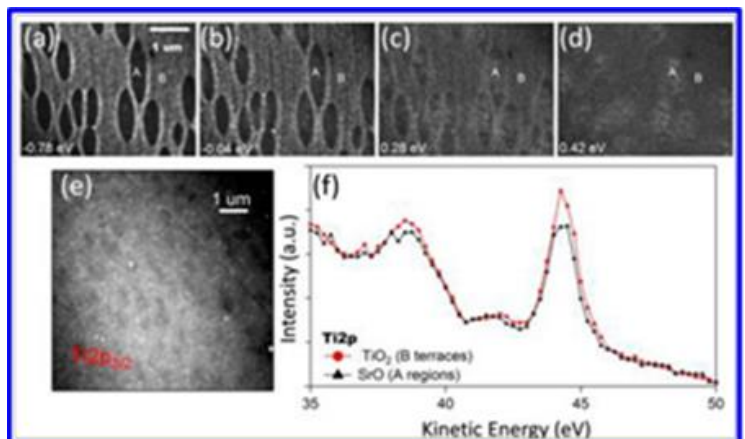

Figure 3. Electron microscopy images across the MEM−LEEM transition (sample 1c). The electron energies are indicated in the lower left corners: (a) −0.78, (b) −0.04, (c) +0.28, and (d) +0.42 eV. (e) XPEEM image at the Ti $2p_{3/2}$ core level, showing lower intensity in the A regions. Photon energy ≈510 eV. (f) Ti 2p core level spectra extracted from both types of surfaces. Notice that the measured kinetic energy is several eV lower than expected, due to sample charging during the photoemission process.

Strong potential gradients, e.g., at steps or at the border between different terminations, cause artifacts such as the bright rims around A regions in Figure 3a (see also Supporting Information), and results in a broadened apparent step width of 50 nm. At first sight one could interpret the observed contrast in terms of differences in local work function. However, MEM image interpretation is tricky with strong potential gradients, especially if charging occurs, as in the present case (see XPEEM cutoff data below) to the point that slight over or underfocusing can result in contrast inversion [26,27]. In summary, though MEM clearly reveals A and B as surface regions with different work function, no quantitative determination can be safely done in this case.

In order to disclose the chemical nature of the uppermost layer in each type of region, we employed X-ray generated photoemission electron microscopy (XPEEM). In XPEEM, the sample is illuminated with photons of known energy, and an image is formed with the emitted photoelectrons. In a spectro- microscope such as the one used here, in addition, the kinetic energy of the electrons can be selected in order to obtain images at any desired region of the spectrum, for example a given core level to obtain elemental maps. We chose the photon energy in order to extract photoelectrons from the Ti 2p core level with a kinetic energy around 50 eV. Such electrons have a very short mean free path or escape depth (≈ 0.5 nm), resulting in extreme surface sensitivity (of the order of 1 uc in this case). With these parameters we acquired X-ray photoemission spectro-microscopy stacks (images at fixed photon energy and varying electron energy) of an area containing both A and B regions. Using them, we extracted spectra from the different regions. The sample is the same as measured previously by nc-AFM (Figure 2e) and MEM (sample 1c), and thus, the shape and relative surface area of both regions are enough to identify A and B regions. An XPEEM image collected at the Ti-$2p_{3/2}$ core level is shown in Figure 3e, revealing a small but very visible contrast between both regions. Photoelectron spectra collected in regions A and B are shown in Figure 3f (see the corresponding Sr 3d spectra in Supporting Information), after proper normalization to the background signal. It can be appreciated that in region B (red circles) the Ti-2p core level peak intensity is larger than in A regions (black triangles), by about 15%. This difference is a consequence of and indication of different Ti content on each surface region. The first and most important result obtained from XPEEM is the unambiguous assignment of terminations A and B to majority SrO and $TiO_2$ compositions, respectively. This photoemission-based assignment is in agreement with contact and noncontact AFM observations including the larger friction measured on A(SrO) with respect to B($TiO_2$), described above. Hence, friction will be from now on the magnitude used for fast identification of majority SrO and $TiO_2$ terminated regions.

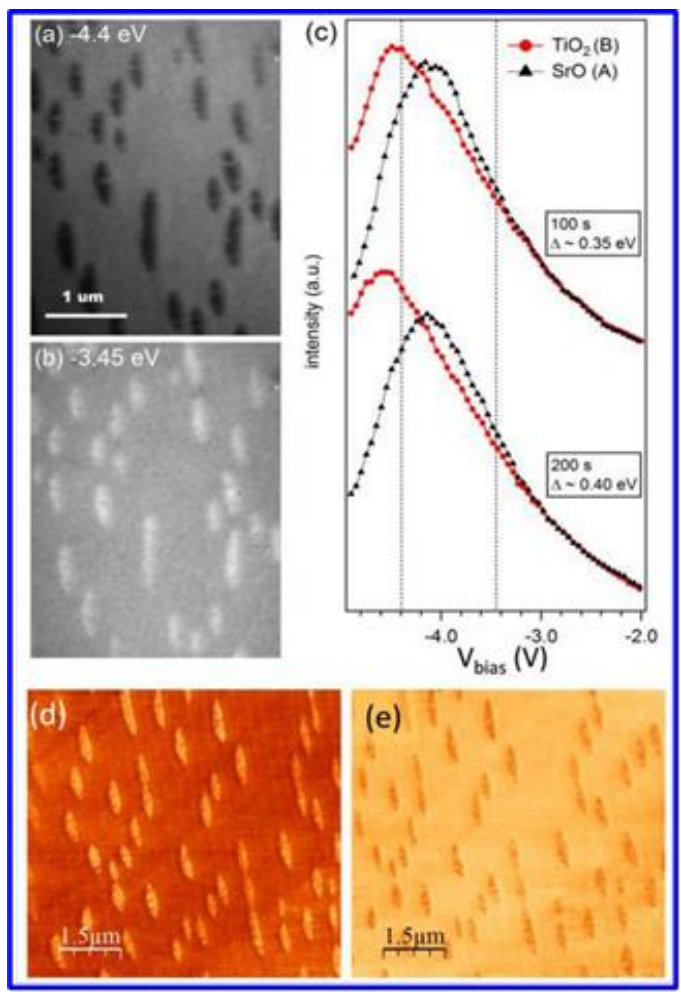

Figure 4. (a, b) XPEEM images (3 μm wide) of sample 1c acquired with 140 eV photon energy at two different bias voltages Vbias, as indicated in the images. (c) Photoemission spectra at the secondary electron cutoff region, extracted from regions A and B for two different measurement speeds: data collection time of 100 s (top pair of spectra) and 200 s (bottom pair). The dashed vertical lines indicate the sample bias at which images a and b were acquired (see text). (d) Forward and (e) backward lateral force images (7.5 μm wide) of the sample obtained by c-AFM in ambient conditions after the XPEEM measurements and therefore corresponding to different surface locations.

Once the majority composition of each region (A = SrO and B = TiO$_2$) was determined, we attempted to determine their work function by measuring the secondary electrons cutoff in XPEEM: the lowest kinetic energy electrons ($E_K$ = 0) of a photoemission spectrum are the most deeply bound electrons that can be extracted from the sample with a given photon energy, e.g., those with $E_{binding}$ = $E_{phot}$ − Φ. Though in conventional spectrometers this part of the spectrum is masked by the work function of the instrument itself, applying a negative bias Vbias to the sample makes it accessible. The XPEEM images of Figure 4a,b, acquired at two different negative bias values, display a bright/dark pattern which mimics that of the FFM images (Figure 4d,e) collected after completion of XPEEM experiments (see below). The work-function-dependent contrast and its inversion at certain Vbias can be understood looking at the low energy cutoff spectra from both regions (Figure 4c). Data obtained from spectro-microscopy stacks taken with different speeds (total time 100 and 200 s for the top and bottom data, respectively) are shown. Surfaces with lower work function have the secondary electrons cutoff at more negative bias, and the intensity relation inverts at a certain point. In both pairs of spectra, the shift between low energy cutoffs of A and B regions indicate a different work function for each termination.

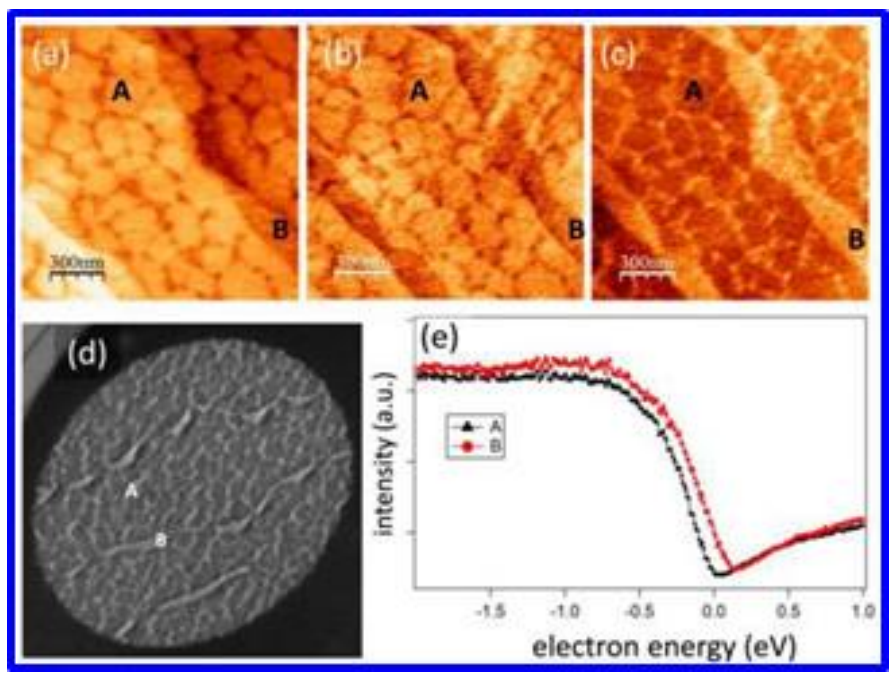

Figure 5. Top: (a) Topography, (b) forward, and (c) backward lateral force images of the as-prepared sample 2a. Lateral force analysis indicates that the surface consists of islands (A) of mainly SrO composition separated by regions (B) of majority TiO2 composition. Bottom: (d) MEM image at −0.06 eV electron energy and (e) MEM−LEEM transition intensity spectra collected at both A and B regions after sample annealing at low temperature (75 °C) in UHV (sample 2b).

On the other hand, it is clear that the cutoff is shifting toward lower values with X-ray irradiation time and the separation between maxima is increasing. In fact, with longer irradiation times (not shown), both cutoffs shift below −5 V, which is the maximum negative sample bias in our instrument [19]. The time- dependent shifts indicate charging of the surface (and more precisely, different charging for both terminations), precluding a precise determination of the work functions or their difference. We can, however, set a sign $\Phi(A) > \Phi(B)$ and an upper limit (≤300 meV) to the work-function difference between A = SrO and B = TiO$_2$ majority terminations. This is in agreement with the CPD data obtained by nc-AFM for this high temperature annealed sample (sample 1c, Figure 2h).

The recurrent question of a possible surface instability and termination intermixing arising from the thermal treatment in reducing conditions was addressed by remonitoring the surface with c-AFM after completion of the XPEEM experiments. Figure 4d,e shows the corresponding forward and backward lateral force images, respectively. Close inspection and comparison between XPEEM and FFM reveals exactly the same surface features, including some lines and spots of inverted contrast visible within the leaf-shaped regions, which are a sign of the local chemical deviations commented when describing Figure 3 (see also FFM analysis in Supporting Information). In summary, in addition to the intrinsic difficulty of obtaining well-separated surface terminations, UHV surface annealing at temperatures as low as ≈400 °C leads to chemical changes which are sufficient to modify the surface potential from that of ideal SrO and TiO$_2$ terminations.

To confirm that the change in sign of the work-function difference observed between as-prepared or mild-annealed samples (samples 1a and 1b) and those annealed at higher temperature (sample 1c) comes from chemical deviations with respect to the nominal SrO and TiO$_2$ stoichiometry due to the thermal treatment employed, we present now results on a different sample (samples 2a and 2b) presenting a different morphology and proportions of A and B regions due to differences on the starting STO crystal. This sample was chosen because it also displays a very recognizable chemical

nano- structuration helpful for data correlation. As before, ambient c- AFM was used to discern by topography and lateral force imaging the spatial distribution of the self-separated terminations on the as-prepared (001)STO (Figure 5a-c). Correlating topography with FFM imaging and knowing that SrO regions present higher friction than the $TiO_2$ ones, we can conclude that this surface has terraces formed by islands of mostly SrO composition (brighter in the forward lateral force image) separated by (mainly) $TiO_2$- terminated elongated and connected regions. A root-mean-square surface roughness rms ~0.15 nm measured on the terraces and height analysis indicate that the islands have the expected ½ uc height of SrO regions on top of the (001)$TiO_2$ plane (see Supporting Information).

In order to minimize any change in surface composition, after introduction in the LEEM−PEEM system, the sample was only annealed at low temperature (75 °C) in UHV (sample 2b), and therefore no surface changes with respect to the as-prepared sample 2a are expected. To obtain reproducible values of the work-function difference, the experimental procedure for determining Φ from the MEM−LEEM transition was optimized. Minimum charging of the surface was achieved by measuring the transition with a very low electron gun intensity and reduced illumination area as well as using an auxiliary UV illumination. Using these conditions stable MEM images could be obtained (Figure 5d). Bias-dependent reflectivity data were acquired at a range of different foci in A and B regions in order to ensure minimal artifacts. As shown in Figure 5e, there is a shift between the MEM−LEEM transition energies (and thus the work function) of both regions, with $\Phi(TiO_2) > \Phi(SrO)$. The obtained difference, $\Phi(TiO_2) - \Phi(SrO) = 70$ meV, is in excellent agreement with the CPD values obtained by KPFM in as-prepared samples (sample 1a, ref 5 and Supporting Information) or by nc-AFM in only slightly annealed samples (sample 1b, Figure 2b) and with the sign expected from calculations for the ideally separated terminations [7].

SUMMARY, CONCLUSIONS, AND OUTLOOK

The (001)$SrTiO_3$ crystal surface can be engineered to display a self-organized structure of nearly pure single-terminated SrO and $TiO_2$ composition by high temperature annealing in oxidizing atmosphere. There are rather well-established protocols for the preparation of STO crystals with controlled surface terminations as reviewed elsewhere [4]. However, the following key points are not solved by conventional tools (e.g., AFM topography) to confirm surface termination: (a) Which is the nature of each termination, that is, SrO or $TiO_2$? (b) Which is the electrostatic potential difference of these surfaces? To address these issues we have selected here samples having different terminations (SrO and $TiO_2$) self-patterned in a clear and recognizable shape to allow exploring different sections of the same surface by using different imaging tools. This is the rationale for using the annealing protocols described in this paper.

By using detailed topographic maps, either by c-AFM and nc- AFM as well as CPD, FFM, and MEM measurements, we have obtained clear evidence of chemical self-structuration of the surfaces of as-prepared crystals. By using XPEEM we have unambiguously identified the surface composition of the self-separated terminations. We have found that these surfaces are extremely sensitive to thermal treatment in reducing conditions, challenging accurate work-function mapping of atomically clean and adsorbate-free chemically nanostructured (001)STO substrates. The sign of the surface potential difference between regions initially consisting of majority single terminations (SrO and $TiO_2$) prepared by annealing crystals in oxygen rich atmosphere at high temperature is found to be $\Phi(SrO) < \Phi(TiO_2)$. These relative values are maintained if samples are annealed, in UHV up to 200 °C, and are in agreement with theoretical predictions, although the measured $\Delta\Phi \leq 100$ meV difference is definitely below the calculated values for pure single-terminated SrO and TiO2 surfaces.

Evidence is found that even mild annealing conditions (200 °C) promote the formation of incipient intermixing. Moreover, although UHV annealing at higher temperature (400 °C) still preserves the surface morphology of self-assembled $TiO_2$ and SrO rich regions, a significant increase of SrO and TiO$_2$ intermixing is observed. The most dramatic consequence is that the surface potential contrast is reversed. It thus follows that electronic and chemical properties of the (001)STO surfaces, widely used in oxide thin film growth, can largely vary before growth starts in a manner strongly depending on the growth conditions (temperature and pressure).

By means of well-defined and controlled examples, this work emphasizes the relevance of full characterization of the surface at the atomistic level of this kind of nanostructured systems, with in-plane modulated electronic properties, which are ideal for serving as substrates for selective growth of thin films with tailored multiferroic, electronic, or catalytic properties. The simplest example of use of these surfaces with coexisting terminations and surface potential modulation could be employing them as templates for the growth of ultrathin (few unit cells) layers of other functional oxides. In fact, it has already been shown that, by using this approach, films (i.e., manganites [7] or ruthenates [6,28]) with laterally modulated friction force and conductivity can be fabricated. Another example could be the use of confined nanometric regions as 2D electron gases3 or local reactors for catalytic (or other) purposes. More visionary possible applications could be lateral p−n junctions or the creation of patterned magnetic regions with different anisotropies (in systems where magnetic anisotropy can be controlled by electric fields and charge density). As reviewed in ref 4, recipes exist to promote chemical self-nanostructuration of STO crystalline surfaces.

ACKNOWLEDGMENTS
This research has been supported by the Spanish MINECO projects: MAT2013-47869-C4-1-P, MAT2011-29269-C03, and MAT2014-56063-C2-1-R. We also acknowledge financial aid from the Generalitat de Catalunya (2014 SGR 501 and SGR 734) and specific agreement between ICMAB-CSIC and the Synchrotron Light Facility ALBA. S.M. thanks the Spanish Government for her BES-2011-045990 FPI fellowship. The LEEM and XPEEM experiments were performed at the CIRCE beamline at ALBA Synchrotron Light Facility with the collaboration of ALBA staff. For their contribution to the CIRCE beamline we especially thank J. Nicolás, E. Pellegrin, S. Ferrer, V. Pérez-Dieste, and C. Escudero. Experimental support of M. Paradinas and L. Garzón acknowledged.